\documentclass[runningheads]{llncs}
\usepackage[T1]{fontenc}
\usepackage{amsmath}
\usepackage{graphicx}
\usepackage{listings}
\usepackage[dvipsnames]{xcolor}
\usepackage{xurl} 
\usepackage{amssymb}
\usepackage{hyperref}
\usepackage{cleveref}

\crefname{figure}{Fig.}{Figs.}   
\Crefname{figure}{Fig.}{Figs.}   
\Crefname{paragraph}{Para.}{Paras.}  
\crefname{paragraph}{Para.}{Paras.}  
\Crefname{definition}{Def.}{Defs.}  
\crefname{definition}{Def.}{Defs.}  

\usepackage{xcolor}
\definecolor{highlight}{RGB}{255,255,150}  
\definecolor{redhighlight}{RGB}{255,180,180}  

\definecolor{yamlkey}{RGB}{0,90,160}       
\definecolor{yamlstr}{RGB}{163,21,21}      
\definecolor{yamlcomment}{RGB}{80,120,70}   
\definecolor{yamlval}{RGB}{110,50,140}      
\definecolor{yamlbg}{RGB}{250,250,248}      

\definecolor{mintbg}{RGB}{248,248,248}      

\usepackage{xspace}
\newcommand{\eg}{\emph{e.g.},\xspace}
\newcommand{\ie}{\emph{i.e.},\xspace}

\newcommand{\cf}{\emph{cf.}\xspace}

\newcommand{\etal}{\emph{et.al.},\xspace}

\newsavebox{\findingboxcontent}

\begin{document}
\pagestyle{headings}
\pagenumbering{gobble}
\title{Where Did the Variability Go? From Vibe Coding to Product Lines by Regeneration}

\author{Xhevahire Tërnava\orcidID{0000-0002-4317-2764}}
\institute{LTCI, Télécom Paris, Institut Polytechnique de Paris, Palaiseau, France \\
\email{xhevahire.ternava@telecom-paris.fr}}

\authorrunning{Xhevahire Tërnava}

\titlerunning{Where Did the Variability Go? VbR Approach}
%
%
%
%
\maketitle              
\begin{abstract}
In vibe coding, an emerging AI-driven paradigm, an LLM generates an entire program from a natural language prompt, but what happens to the variability that traditional software engineering carefully builds into code?
To answer this question, we conducted an exploratory analysis on 10 vibe coded C/C++ projects, which suggests that there is near zero in-artifact variability, \ie at compile- and runtime. 
All variability decisions are resolved at a single new binding time, \emph{generation time}, the moment the LLM produces the source code.
Rather than treating this as a defect to fix, we propose \emph{Variability by Regeneration}~(VbR), to our knowledge the first product line approach in which the LLM acts as the derivation engine, generating a dedicated, free of dead code binary for each variant from a declarative specification, while a variant dispatcher transparently routes user requests to the matching binary.
We formalise VbR, contrast it with classical SPL derivation, and demonstrate its full pipeline on a \texttt{wc} product family. For SPL engineering, variability in AI-generated software belongs in the specification, not in the code.
\end{abstract}
%

\section{Introduction}
\label{introduction}
The ability to configure, extend, and customise a software system for different contexts, what is known in the literature as \emph{software variability}, is a central concern of software engineering~\cite{clements2001software}. Over four decades, the field has built a rich set of approaches and tools to implement and manage variability systematically, \eg feature models~\cite{kang1990feature}, variation points, as embedded choices in code~\cite{czarnecki2012cool}, and product derivation mechanisms~\cite{pohl2005software}. Across these techniques, one principle has been constant, namely variability should be \emph{planned} and \emph{in-artifact}, realised through explicit mechanisms that make configuration and evolution tractable.

A recent shift in how software is produced is challenging this principle.
In February 2025, Karpathy coined the term ``vibe coding'' to describe an ongoing development approach in which users describe their intent in natural language and delegate implementation entirely to an AI tool, such as Cursor, Claude, or Copilot~\cite{karpathy2025vibe}.
The approach has spread rapidly, motivated by speed and accessibility~\cite{fawzy2025vibe}, reframing the developer's role from implementer to intent mediator~\cite{meske2025vibe}.
Yet, \textit{no study has examined what happens to variability when a system is vibe coded, from the perspective of software product line (SPL) engineering}.

Our starting observation, detailed in \Cref{backgroundandmotivation}, is that vibe coded systems tend to exhibit near zero in-artifact variability at every binding time. All variability decisions are effectively frozen at what we call \emph{generation time}, \ie the moment the LLM produces the source code.
Rather than viewing this absence of variability as a problem to fix, for instance by teaching LLMs to emit \texttt{\#ifdef} guards, we propose \emph{Variability by Regeneration}~(VbR), an approach that embraces zero-variability code~\cite{ternava2025nullvariability,acher2023removing} as the intended output and manages variability entirely outside the code artefact.
In VbR, the developer declares a product family, \ie features, variant configurations, and binding times, in a single \emph{specification}. Then, an LLM \emph{generates} a dedicated 
binary for each software variant, and a \emph{dispatcher} transparently routes user requests to the matching binary.

This paper contributes VbR, a formally defined product line approach that, to our knowledge, is the first designed for AI-generated software, using the LLM as the derivation engine and introducing \emph{generation time} as a new binding time, together with a replication package~\footnote{Replication package: \url{https://doi.org/10.5281/zenodo.20730698}} containing the complete application pipeline of VbR on the \texttt{wc} product family, and the data of our exploratory analysis.
VbR is admittedly a somewhat controversial proposal, but we believe it is worth considering as vibe coding reshapes how software is produced, and we are explicit about the limitations (\cf~\Cref{limitations}) that currently bound it.

\Cref{backgroundandmotivation} provides background and motivation, \Cref{sec:vbr} introduces and formalises VbR, \Cref{relatedwork} presents related work, and \Cref{conclusion} concludes.

\section{Background and Motivation}
\label{backgroundandmotivation}

\paragraph{Vibe Coding as a Paradigm Shift.}
While most AI-assisted development keeps the human in the authoring loop, \emph{vibe coding}~\cite{karpathy2025vibe} is qualitatively different. The LLM \emph{replaces} the entire act of programming, as the user simply describes a system in natural language and receives a complete, runnable program.
Since early 2025, thousands of self-identified vibe coded repositories have appeared on GitHub~\cite{fawzy2025vibe,github2025octoverse01}, and empirical studies report that practitioners treat the resulting code as disposable, to be accepted or regenerated rather than maintained~\cite{chou2025building,pimenova2025good}. Although only about 15\% of professional developers currently practise vibe coding~\cite{stackoverflow2025survey}, the code it produces is already deployed as real software~\cite{fawzy2025vibe,superwhisper2025,yang2025vibefps}.

\paragraph{The Variability Gap.}
A growing body of work assesses the quality of AI-generated code, identifying overlooked reuse opportunities~\cite{huang2026more}, unnecessary functions~\cite{watanabe2026cut}, elevated defect rates~\cite{cotroneo2025human}, and hidden technical debt~\cite{mikkonen2025reuse,pimenova2025good}, but always at the granularity of individual programs, functions, or files.
None has examined a property that is fundamental at the \emph{system} level, namely \emph{software variability}, realised in traditionally engineered systems through explicit mechanisms at well-defined binding times (\eg preprocessor directives at compile-time, component substitution at link-time, CLI options at runtime~\cite{apel2013feature,capilla2013binding,svahnberg2005taxonomy,ternava2017diversity}), the very mechanisms that transform an individual program into a product family~\cite{clements2001software}.
If vibe coded systems lack them entirely, then every generated program is a single-purpose artefact and adapting it means regenerating from scratch, \ie clone-and-own at industrial scale, the very anti-pattern SPL engineering was designed to prevent.

\paragraph{A Motivating Observation.}
This paper thus starts from a question not posed before: \emph{does vibe coded software retain any in-artifact variability, or does the act of generation eliminate it?}
To probe it, we conducted an exploratory analysis of 10~vibe coded C/C++ projects from GitHub, spanning seven domains and two orders of magnitude in size (201 to 14{,}077~LoC), compared against \texttt{x264}, a long-lived configurable C system of 70{,}964~LoC~\cite{lesoil2021interaction,merritt2006x264,ternava2023specialization}.
The contrast is evident. While \texttt{x264} exposes 184~CLI options and references 117~distinct preprocessor variables across 2{,}201 conditional compilation directives, the 10 vibe coded projects collectively expose only 45~CLI options (median~0, with 6 of 10 exposing none) and reference just 27~preprocessor variables, none of which represents \emph{designed} variability (all are incidental platform checks, math constant fallbacks, or vendored library internals).
Nor does project size explain the gap, \eg GNU~\texttt{wc}, roughly 800~LoC, exposes 8~CLI options and multiple conditional compilation paths~\cite{gnu-coreutils-2026,luu2020cli}, despite being smaller than 80\% of our 10 vibe coded subjects.
This whole exploratory analysis is available in our replication package.

\paragraph{Generation Time: A New Binding Time.}
Traditionally, variability decisions are resolved progressively along a binding-time pipeline, from feature selection at specification time through compilation and linking to CLI options at runtime, and the delivered product \emph{retains} runtime variability~\cite{capilla2013binding}.
Based on our exploratory analysis, this pipeline collapses into a single moment we call \emph{generation time}, at which, unless the developer specifies otherwise, the LLM resolves all variability decisions, yielding a fixed binary, for a single purpose, with no variability mechanisms (\eg no \texttt{\#ifdef} and no \texttt{getopt()}).
Understanding what this means for engineering AI-generated software that practitioners can configure, reuse, and extend is the central goal of this work.

\section{Variability by Regeneration (VbR)}
\label{sec:vbr}

\subsection{Regenerate, Do Not Configure}
\label{sec:VbRapproach}
As motivated in \Cref{backgroundandmotivation}, vibe coded systems tend to have zero variability in the code artifact. We embrace this and manage variability entirely outside the codebase.
In traditional software engineering, variability is expensive to implement but cheap to exploit, \ie an organisation invests upfront effort building a shared asset base with variation points~\cite{apel2013feature,clements2001software}, then derives tailored products cheaply from it.
AI code generation inverts this cost structure, where generation is cheap (from the developer's perspective), but the resulting code is already hard for humans to understand and maintain~\cite{cotroneo2025human}, and layering variation points on top makes it even harder.
Instead of generating one configurable system, we therefore propose \emph{generating many software systems, each corresponding to a single variant, with zero internal variability~\cite{ternava2025nullvariability}, and dispatch the one that fits the user's current context}, an approach we call \emph{Variability by Regeneration}~(VbR).
In VbR, the core artefact of the product family is a \emph{feature specification}, a \emph{generation pipeline}, and a \emph{dispatch mechanism}, so each variant is a freshly generated, dedicated artefact containing only the code paths it needs.
This design aligns with the emerging practice where tools such as Claude Code and Cursor treat persistent specification files (\texttt{CLAUDE.md}, \texttt{.cursor/rules}) as the primary artefact driving generation~\cite{anthropic2025claudemd,cursor2025rules}. Our VbR approach generalises this idea from rules for a single product to a specification for a family of software variants.

\subsection{A Formal Description of VbR}
\label{sec:vbr-formal}
VbR consists of three components: \emph{(i)}~a \emph{specification} $\mathcal{S}$ declaring features, their types and constraints, and the variant configurations; \emph{(ii)}~a \emph{generation function} $G$ synthesising a code artefact per variant; and \emph{(iii)}~a \emph{variant dispatcher} $D$ routing user requests to the appropriate variant. 

\begin{definition}[VbR Specification]\label{def:vbr-spec}
A \emph{VbR product family} is defined by a specification $\mathcal{S} = \langle F, \tau, \mathcal{V}, \mathcal{C} \rangle$, where
$F = \{f_1, \ldots, f_n\}$ is a finite set of \emph{features};
$\tau : F \to \{\textit{bool}, \textit{enum}, \textit{int}, \textit{string}\}$ assigns a type to each feature;
$\mathcal{V} = \{v_1, \ldots, v_m\}$ is a finite set of \emph{variant configurations}; and
$\mathcal{C} \subseteq F \times \{\textit{requires}, \textit{excludes}\} \times F$ is a set of cross-feature constraints.
Each \emph{variant configuration} $v_j \in \mathcal{V}$ is a partial function $v_j : F \rightharpoonup \mathit{Val}$, where $\mathit{Val}$ is the union of all value domains of~$\tau$, and each included feature is mapped to a value. Features absent from $\mathrm{dom}(v_j)$ are excluded and generate no code.
A configuration $v_j$ is \emph{valid} if it is \emph{(1)~well typed}, \emph{(2)~closed under constraints}, \ie if $f_a$ requires $f_b$ and $f_a \in \mathrm{dom}(v_j)$, then $f_b \in \mathrm{dom}(v_j)$, and \emph{(3)~free of conflicts}, \ie if $f_a$ excludes $f_b$, they cannot both be in $\mathrm{dom}(v_j)$.
\end{definition}

\begin{definition}[Generation Function]\label{def:vbr-gen}
Beyond the specification $\mathcal{S}$, VbR relies on two additional inputs: a \emph{prompt template} $\Pi$ (encoding generation rules, coding conventions, and an annotation format for traceability) and a \emph{language model} $\mathfrak{L}$ (\ie LLM).
A generation function is a mapping $G : \mathcal{S} \times \mathcal{V} \times \Pi \times \mathfrak{L} \to A$ that takes a specification, a valid configuration $v_j$, a prompt template, and a language model, and produces a code artefact (a compilable software variant) $A_j$.
\end{definition}
Note that $v_j$ is a variant configuration, while $A_j = G(\mathcal{S}, v_j, \Pi, \mathfrak{L})$ is the code artefact generated from it, equipped with a \emph{traceability relation} $\mathcal{T}_j \subseteq F \times \mathrm{Loc}(A_j)$ recording which source locations implement each included feature.
Unlike classical variant derivation, the generation function $G$ takes \emph{no pre-existing code asset base} as input, \ie in the VbR approach, a variant is synthesised, not selected.

\begin{definition}[Variant Dispatcher]\label{def:vbr-dispatch}
Given the code artefacts $\{A_1, \ldots, A_m\}$ generated from the variant configurations in $\mathcal{S}$, a \emph{variant dispatcher} is a pair $D = \langle \mathcal{M}, \delta \rangle$, where
$\mathcal{M} : \mathcal{V} \to \mathrm{Caps} \times \mathrm{Path}$ is a \emph{manifest} mapping each variant configuration to its capability set, \ie the features in $\mathrm{dom}(v_j)$, and its binary path, whereas
$\delta : \mathrm{Request} \to \mathcal{V}$ is a \emph{dispatch function} that maps a user's feature request to the realized variant (program) whose capabilities best match the request.
\end{definition}
By default, each artefact $A_j$ has zero internal variability (Prop.~\ref{prop:zdc}), \ie it exposes \emph{no configuration surface}. Consequently, adapting the system means selecting a \emph{different} artefact, so dispatch operates \emph{between programs}, not within a program.

The following 3 properties are \emph{generation pipeline contracts}, not guarantees the LLM provides on its own. The pipeline checks each one on every artefact $A_j$, via static checks over the traceability relation $\mathcal{T}_j$ and a gate that compiles and tests the artefact. Any artefact violating a contract is rejected and regenerated, so the properties hold by construction of the pipeline, not by trust in the model.

\begin{property}[Zero Dead Code]\label{prop:zdc}
For every valid configuration $v_j$ and feature $f \in F$, if $f$ is not included in $v_j$, then it is not present in  $A_j$, \ie $f \notin \mathrm{dom}(v_j) \implies \nexists\, \ell \in \mathrm{Loc}(A_j) : (f, \ell) \in \mathcal{T}_j$.
That is, excluded features produce \emph{no} code in the generated artefact $A_j$, not even dead code behind unreachable guards.
\end{property}

\begin{property}[Traceability Completeness]\label{prop:tc}
For any valid configuration $v_j$ and included feature $f \in \mathrm{dom}(v_j)$, there must be at least one source code location in $A_j$ that implements $f$:
$f \in \mathrm{dom}(v_j) \implies \exists\, \ell \in \mathrm{Loc}(A_j) : (f, \ell) \in \mathcal{T}_j$.
Together with Prop.~\ref{prop:zdc}, this guarantees a bijective correspondence between specification and code. Every specified feature $f$ is in the code, and nothing in the code is without a specified $f$, a mapping that both the dispatcher and evolution rely on.
\end{property}

\begin{property}[Binding Time Flexibility]\label{prop:btf}
Let $\mathit{BT} = \{gen, compile, link, \ldots, rt\}$ be the set of \emph{binding times}, from generation time to runtime.
Each variant configuration $v_j$ may be extended with a binding annotation, that assigns a binding time to each included feature, \ie $\beta_j : \mathrm{dom}(v_j) \to \mathit{BT}$. When $\beta_j$ is omitted, the default $\beta_j(f) = gen$ applies, \ie the variant is a fixed product free of variability. Then, the same feature~$f$ may have $\beta_j(f) = gen$ in one variant and $\beta_k(f) = rt$ in another.
In classical SPL, binding time is a \emph{consequence} of the chosen implementation mechanism, whereas in VbR it is a \emph{design decision per feature per variant}, declared in the specification $\mathcal{S}$ and enforced by the generator $G$.
\end{property}

\begin{figure}[t]
\centering
\includegraphics[width=\textwidth]{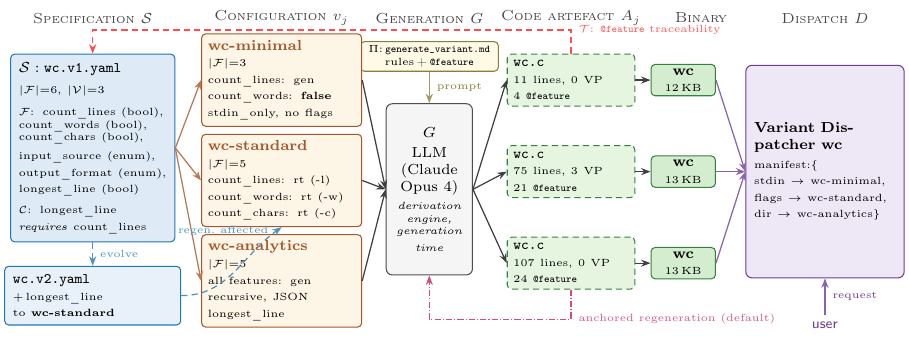}
\caption{The VbR pipeline instantiated for the \texttt{wc} product family.}
\label{fig:vbr}
\end{figure}

\paragraph{Contrast with Classical SPL.}
In the standard SPL framework, a product line is a triple $(\mathit{FM}, \mathcal{A}, d)$, \ie a feature model $\mathit{FM}$ defining valid configurations, an \emph{asset base} $\mathcal{A}$ of implementation artefacts with embedded variation points, and a derivation function $d : 2^F \times \mathcal{A} \to P$ that selects and composes artefacts into a product.
Crucially, $\mathcal{A}$ \emph{pre-exists}, \ie every variation point and every feature interaction is encoded in the asset base before any product is derived.
VbR, in contrast, has \emph{no asset base}. The specification $\mathcal{S}$ is the only persistent artefact, and code is a regenerated by-product of $G$.
This is why we deliberately call the result a \emph{product family} rather than a product line. Products are not derived from a shared asset base, with no guarantee of code-level commonality across variants.
In short, classical SPL maintains code and derives products by selection, whereas VbR maintains only the specification and derives products by synthesis.
\Cref{fig:vbr} illustrates the VbR pipeline end-to-end, instantiated next on a concrete example.

\subsection{Instantiating VbR: The \texttt{wc} Product Family}
\label{sec:vbr-pipeline}
To demonstrate VbR, we use the Unix \texttt{wc} (word count) utility~\cite{gnu-coreutils-2026}, simple to show in full yet rich on variability. Its implementation is in the replication package.\footnote{The \href{https://doi.org/10.5281/zenodo.20730698}{\texttt{vbr-demo/}} directory of the replication package.}

\paragraph{Specification.}
\label{sec:vbr-example}
The top of~\Cref{fig:vbr} shows the specification $\mathcal{S}$ (\Cref{def:vbr-spec}), a single YAML file (\texttt{wc.v1.yaml}) declaring the feature space $F$, constraints $\mathcal{C}$, and all variant configurations $\mathcal{V}$.
The \texttt{wc} family has six features, four booleans and two enums, with one constraint (\texttt{longest\_line} \emph{requires} \texttt{count\_lines}). \texttt{wc.v1.yaml} and the prompt template \texttt{generate\_variant.md} are the only two files written by hand.

\paragraph{Configuration and Binding Times.} 
\label{para:config-binding}
The three variant configurations are defined in the same YAML file, each as a partial function $v_j$ (\Cref{def:vbr-spec}) that selects feature values and assigns binding times. The configuration boxes in \Cref{fig:vbr} show the key design decisions.
By convention, every feature, included or not, is explicitly listed per variant (Prop.~\ref{prop:zdc}). Hence, an excluded feature is set to \texttt{false} (\eg \texttt{count\_words: false} in \texttt{wc-minimal}), signalling the generator to produce no code for it. As for the feature constraints, they are checked by a YAML-parsing script \emph{before} the LLM is invoked.
Binding times are chosen per variant (Prop.~\ref{prop:btf}). For instance, the \texttt{count\_lines} in \texttt{wc-standard} is bound at runtime ($b = rt$) and exposed as \texttt{-l}, whereas in \texttt{wc-minimal} it is bound at generation time ($b = gen$), meaning that it is always enabled, requiring no flag parser.

\paragraph{Generation.}
For each variant, 
the generation function $G$ (\Cref{def:vbr-gen}) feeds the specification $\mathcal{S}$ (\texttt{wc.v1.yaml}), the configuration~$v_j$ (\eg \texttt{wc-minimal}), and the prompt template $\Pi$ (\texttt{generate\_variant.md}) to an LLM (Claude Opus~4 here) and obtains a self-contained \texttt{wc.c} file per variant. 
Each variant includes only the features its configuration specifies, with no dead code, where binding is at generation time, and runtime variation points only where late binding is requested (\eg \texttt{wc-standard}).
The generator also emits \texttt{@feature} annotations tracing every code element back to its feature (Prop.~\ref{prop:tc}), shown as dashed arrows in~\Cref{fig:vbr}.

\paragraph{Variant Dispatch.}
After generation and compilation, the family consists of three standalone binaries, which the variant dispatcher $D = \langle \mathcal{M}, \delta \rangle$ (\Cref{def:vbr-dispatch}) exposes as a single \texttt{wc} command. Its two components are specific to the \texttt{wc} family: the manifest $\mathcal{M}$ (\texttt{manifest.yaml}) which encodes dispatch rules, ordered by priority, mapping usage patterns to binaries, and the dispatch function $\delta$, a shell script (\texttt{dispatch.sh}), parses the user's command line, evaluates the rules top-down, and \texttt{exec()}s the matching binary. For example:
\begin{lstlisting}[language=bash,basicstyle=\ttfamily\footnotesize,backgroundcolor=\color{cyan!10},frame=none,numbers=none,columns=fixed,showstringspaces=false,breaklines=true]
$echo "hello world" | wc      # dispatches to wc-minimal 
$wc -l -w report.txt          # dispatches to wc-standard
\end{lstlisting}
\noindent The user never names a variant. The dispatcher infers it from the flags and input mode.
This inverts BusyBox's~\cite{busybox} bundling of many utilities into one binary.

\subsection{Discussion: Applicability, Evolution, and Limitations}
\label{sec:vbr-discussion}

VbR targets the developer profile that vibe coding attracts~\cite{fawzy2025vibe,karpathy2025vibe}, \ie practitioners who generate entire programs from prompts, bridging this workflow with the disciplined variability management a product family requires~\cite{apel2013feature,pohl2005software}.

\paragraph{Evolution Through Specification Change.}
Adding or removing a feature is a one line change in the specification, in which case only the affected variant will be regenerated.
In our demonstration, enabling \texttt{longest\_line} in \texttt{wc-standard} required a single edit to \texttt{wc.v1.yaml} (blue dashed arrow in \Cref{fig:vbr}), yielding \texttt{wc.v2.yaml}.
Prop.~\ref{prop:zdc} guarantees the regenerated variant contains zero dead code and the \texttt{@feature} annotations (Prop.~\ref{prop:tc}) identify which code locations a change affects. Moreover, the specification, versioned in a VCS, serves as the evolution log, and a compilation and test gate, with per-feature acceptance tests optionally declared in $\mathcal{S}$, checks each regeneration for regressions.
For larger variants, VbR also envisions an \emph{anchored regeneration} mode (purple dash-dot arrows in \Cref{fig:vbr}), where the existing variant code plus the specification diff yield a targeted update rather than a full rewrite. When the specification is pinned to a version and the model is held fixed, this mode also handles bug fixing, since a fix becomes a localized and reproducible regeneration of the affected variant rather than a fresh synthesis, so unrelated code does not drift. This mode of VbR was not exercised for the \texttt{wc} variants given their small size.

\paragraph{Limitations.}
\label{limitations}
VbR inherits the limitations of its generator ($G$). Namely, LLM output is non-deterministic and may introduce subtle bugs or insecure idioms~\cite{cotroneo2025human}, and it offers no formal correctness guarantee. Prompt safety rules and post-generation analysis and test gates mitigate but do not eliminate this.
Secondly, VbR sacrifices code reuse across variants, as two variants including the same feature may receive different implementations~\cite{mikkonen2025reuse}, a deliberate trade-off for smaller, focused binaries, but a bug fix in one variant does not automatically propagate.
Thirdly, our evaluation is limited to one small product family. Scaling to tens of variants, and measuring generation cost, LLM consistency, and dispatcher overhead, remain open.
Finally, verification of the three VbR properties is currently manual. Automating it in the pipeline is necessary for production use.

\section{Related Work}
\label{relatedwork}
Variability is a central concern in SPL engineering~\cite{apel2013feature,pohl2005software}, studied through many aspects, such as preprocessor usage in C/C++ systems~\cite{liebig2010analysis}, industrial modelling practices~\cite{berger2015feature}, binding-time taxonomies~\cite{capilla2013binding,czarnecki2000generative,svahnberg2005taxonomy}, and traceability techniques~\cite{ji2015techniques,ternava2017tracing}. Whether AI-generated code exhibits these patterns has, to our knowledge, not been systematically studied, as we did in the exploratory study.

LLM-generated code has been studied \eg for functional quality~\cite{chen2021evaluating}, security vulnerabilities~\cite{pearce2022asleep}, and recurring bugs~\cite{jesse2023large}. These works address correctness, whereas we examine \emph{architectural variability}, a design level property they do not capture.
On the other hand, vibe coding~\cite{karpathy2025vibe} has attracted growing empirical investigation~\cite{chou2025building,fawzy2025vibe,meske2025vibe,pimenova2025good}. We complement these works by examining its variability implications and responding with a constructive, product line approach.

Acher~\etal~\cite{acher2023programming} showed that LLM-based assistants help implement variability mechanisms, whereas VbR removes them altogether.
St\"{u}mpfle~\etal~\cite{stumpfle2025llm} and Acher ~\etal~\cite{acher2023reengineering} used LLMs to re-engineer cloned variants into an SPL, the reverse direction of VbR. Both assume a single variable code base as the end-state.
Greiner~\etal~\cite{greiner2024vision} envisioned GAI-assisted automation in variability-intensive systems. VbR is a concrete instantiation of it, but eliminates variability mechanisms rather than automating their management.
Zine~\etal~\cite{zine2025coevolution} applied LLMs to co-evolve configurable systems when features change. In VbR, 
this means updating the specification and regenerating.
Sonkin and Tudose~\cite{sonkin2025workflow} proposed regeneration-and-rollback workflows, but for single program automation with no notion of features. In contrast, VbR generates systems with zero variability where each variant contains exactly the code it needs, managing the family through a shared specification and serving each variant through a dispatcher.

\section{Conclusion and Future Work}
\label{conclusion}
Vibe coding moves variability out of the code and into the prompt.
Rather than treating this as a problem to fix, we proposed and formalised \emph{Variability by Regeneration}~(VbR), to our knowledge the first product line approach designed for AI-generated software. In VbR, variability lives in a declarative specification, and each variant is a freshly generated, dedicated binary with no dead code (\ie no variability branches) and full traceability, as we demonstrated on a real (\texttt{wc}) product family with six features and three variants.
Central to VbR is \emph{generation time}, the new binding time at which the LLM resolves every variability decision before any source code exists, a shift our exploratory analysis of 10~vibe coded projects suggests is already common in practice.

\paragraph{Future Work.} 
Our immediate next step is an empirical study of in-artifact variability in vibe coded systems, across more languages and domains. We then plan to scale VbR to tens of variants, automate property verification, and develop anchored regeneration to cut generation cost and support in-artifact reuse.

As vibe coding becomes more common, variability moves from the artefact to the specification. VbR is a first step towards PLs built entirely on that shift.

%
%
\bibliographystyle{splncs04}
\bibliography{vibe_coding_refs}

\begin{thebibliography}{10}
\providecommand{\url}[1]{\texttt{#1}}
\providecommand{\urlprefix}{URL }
\providecommand{\doi}[1]{https://doi.org/#1}

\bibitem{acher2023programming}
Acher, M., Duarte, J.A.G., J{\'{e}}z{\'{e}}quel, J.: On programming variability
  with large language model-based assistant. In: Proc.\ {SPLC}. pp. 8--14.
  {ACM} (2023). \doi{10.1145/3579027.3608972}

\bibitem{acher2023removing}
Acher, M., Lesoil, L., Randrianaina, G.A., T{\"e}rnava, X., Zendra, O.: A call
  for removing variability. In: Proceedings of the 17th International Working
  Conference on Variability Modelling of Software-Intensive Systems (VaMoS).
  {ACM} (2023). \doi{10.1145/3571788.3571801}

\bibitem{acher2023reengineering}
Acher, M., Martinez, J.: Generative {AI} for reengineering variants into
  software product lines: An experience report. In: Proc.\ {SPLC}. pp. 57--66.
  {ACM} (2023). \doi{10.1145/3579028.3609016}

\bibitem{anthropic2025claudemd}
{Anthropic}: How {Claude} remembers your project: {CLAUDE.md} files. Claude
  Code Documentation (2025), \url{https://code.claude.com/docs/en/memory}

\bibitem{apel2013feature}
Apel, S., Batory, D., K{\"a}stner, C., Saake, G.: Feature-Oriented Software
  Product Lines: Concepts and Implementation. Springer (2013).
  \doi{10.1007/978-3-642-37521-7}

\bibitem{berger2015feature}
Berger, T., Nair, D., Rublack, R., Atlee, J.M., Czarnecki, K., W{\k{a}}sowski,
  A.: Three cases of feature-based variability modeling in industry. pp.
  302--319. Springer (2014). \doi{https://doi.org/10.1007/978-3-319-11653-2_19}

\bibitem{busybox}
{BusyBox}: {BusyBox}: The {Swiss} army knife of embedded {Linux}. Online
  software (2024), \url{https://busybox.net/}

\bibitem{capilla2013binding}
Capilla, R., Bosch, J.: Binding time and evolution. In: Capilla, R., Bosch, J.,
  Kang, K.C. (eds.) Systems and Software Variability Management: Concepts,
  Tools and Experiences, pp. 57--73. Springer (2013).
  \doi{10.1007/978-3-642-36583-6_4}

\bibitem{chen2021evaluating}
Chen, M., Tworek, J., Jun, H., Yuan, Q., de~Oliveira~Pinto, H.P., Kaplan, J.,
  Edwards, H., Burda, Y., Joseph, N., Brockman, G., et~al.: Evaluating large
  language models trained on code (2021). \doi{10.48550/arXiv.2107.03374}

\bibitem{chou2025building}
Chou, Y., Jiang, B., Chen, Y.W., Weng, M., Jackson, V., Zimmermann, T., Jones,
  J.A.: Building software by rolling the dice: {A} qualitative study of vibe
  coding. In: Proc.\ {ESEC/FSE} (2026). \doi{10.48550/arXiv.2512.22418}

\bibitem{clements2001software}
Clements, P., Northrop, L.: Software Product Lines: Practices and Patterns.
  Addison-Wesley (2001)

\bibitem{cotroneo2025human}
Cotroneo, D., Improta, C., Liguori, P.: Human-written vs.\ {AI}-generated code:
  {A} large-scale study of defects, vulnerabilities, and complexity. In: Proc.\
  {ISSRE}. pp. 252--263. {IEEE} (2025). \doi{10.1109/ISSRE66568.2025.00035}

\bibitem{cursor2025rules}
{Cursor Inc.}: Rules: system-level instructions for {Agent}. Cursor
  Documentation (2025), \url{https://cursor.com/docs/rules}

\bibitem{czarnecki2000generative}
Czarnecki, K., Eisenecker, U.W.: Generative Programming: Methods, Tools, and
  Applications. Addison-Wesley (2000)

\bibitem{czarnecki2012cool}
Czarnecki, K., Gr{\"u}nbacher, P., Rabiser, R., Schmid, K., W{\k{a}}sowski, A.:
  Cool features and tough decisions: {A} comparison of variability modeling
  approaches. In: Proc.\ 6th International Workshop on Variability Modelling of
  Software-Intensive Systems ({VaMoS}). pp. 173--182. {ACM} (2012).
  \doi{10.1145/2110147.2110167}

\bibitem{fawzy2025vibe}
Fawzy, A., Tahir, A., Blincoe, K.: Vibe coding in practice: Motivations,
  challenges, and a future outlook--a grey literature review (2025).
  \doi{10.48550/arXiv.2510.00328}

\bibitem{gnu-coreutils-2026}
{Free Software Foundation}: {GNU} Coreutils: Core {GNU} utilities. Free
  Software Foundation (2026), \url{https://www.gnu.org/software/coreutils/},
  version 9.10

\bibitem{github2025octoverse01}
{GitHub}: Octoverse: A new developer joins {G}it{H}ub every second as {AI}
  leads {T}ype{S}crip to\#1 (2025),
  \url{https://github.blog/news-insights/octoverse/octoverse-a-new-developer-joins-github-every-second-as-ai-leads-typescript-to-1/},
  accessed: 2026-07-08. September 2024--August 2025 data

\bibitem{greiner2024vision}
Greiner, S., Schmid, K., Berger, T., Krieter, S., Meixner, K.: Generative {AI}
  and software variability--{A} research vision. In: Proc.\ {VaMoS}. pp.
  71--76. {ACM} (2024). \doi{10.1145/3634713.3634722}

\bibitem{huang2026more}
Huang, H., Jaisri, P., Shimizu, S., Chen, L., Nakashima, S.,
  Rodr{\'\i}guez{-}P{\'e}rez, G.: More code, less reuse: Investigating code
  quality and reviewer sentiment towards {AI}-generated pull requests. In:
  Proc.\ {MSR} (2026). \doi{10.48550/arXiv.2601.21276}, to appear

\bibitem{jesse2023large}
Jesse, K., Ahmed, T., Devanbu, P.T., Morgan, E.: Large language models and
  simple, stupid bugs. In: Proc.\ {MSR}. pp. 563--575. {IEEE} (2023).
  \doi{10.1109/MSR59073.2023.00082}

\bibitem{ji2015techniques}
Ji, W., Berger, T., Antkiewicz, M., Czarnecki, K.: Maintaining feature
  traceability with embedded annotations. In: Proc.\ {SPLC}. pp. 61--70. {ACM}
  (2015). \doi{10.1145/2791060.2791107}

\bibitem{kang1990feature}
Kang, K.C., Cohen, S.G., Hess, J.A., Novak, W.E., Peterson, A.S.:
  Feature-oriented domain analysis ({FODA}) feasibility study. Tech. Rep.
  CMU/SEI-90-TR-021, Software Engineering Institute (1990).
  \doi{10.21236/ADA235785}

\bibitem{karpathy2025vibe}
Karpathy, A.: Vibe coding. X (Twitter) (Feb 2025),
  \url{https://x.com/karpathy/status/1886192184808149383}

\bibitem{lesoil2021interaction}
Lesoil, L., Acher, M., Blouin, A., J{\'e}z{\'e}quel, J.: The interaction
  between inputs and configurations fed to software systems: An empirical
  study. CoRR  \textbf{abs/2112.07279} (2021). \doi{10.48550/arXiv.2112.07279},
  journal reference: Journal of Systems and Software, 2023

\bibitem{liebig2010analysis}
Liebig, J., Apel, S., Lengauer, C., K{\"a}stner, C., Schulze, M.: An analysis
  of the variability in forty preprocessor-based software product lines. In:
  Proc.\ {ICSE}. pp. 105--114 (2010). \doi{10.1145/1806799.1806819}

\bibitem{luu2020cli}
Luu, D.: The growth of command line options, 1979--present.
  \url{https://danluu.com/cli-complexity/} (2020)

\bibitem{merritt2006x264}
Merritt, L., Vanam, R.: x264: {A} high performance {H.264/AVC} encoder. Online
  software (2006), \url{https://code.videolan.org/videolan/x264}

\bibitem{meske2025vibe}
Meske, C., Hermanns, T., von~der Weiden, E., Loser, K., Berger, T.: Vibe coding
  as a reconfiguration of intent mediation in software development: Definition,
  implications, and research agenda. IEEE Access  \textbf{13},  213242--213259
  (2025). \doi{10.1109/ACCESS.2025.3645466}

\bibitem{mikkonen2025reuse}
Mikkonen, T., Taivalsaari, A.: Software reuse in the generative {AI} era: From
  cargo cult towards systematic practices. In: Proc.\ Internetware. pp.
  541--544. {ACM} (2025). \doi{10.1145/3755881.3755981}

\bibitem{pearce2022asleep}
Pearce, H., Ahmad, B., Tan, B., Dolan{-}Gavitt, B., Karri, R.: Asleep at the
  keyboard? {Assessing} the security of {GitHub Copilot}'s code contributions.
  In: Proc.\ {IEEE} Symposium on Security and Privacy ({SP}). pp. 754--768.
  {IEEE} (2022). \doi{10.1109/SP46214.2022.9833571}

\bibitem{pimenova2025good}
Pimenova, V., Fakhoury, S., Bird, C., Storey, M., Endres, M.: Good vibrations?
  {A} qualitative study of co-creation, communication, flow, and trust in vibe
  coding (2025). \doi{10.48550/arXiv.2509.12491}

\bibitem{pohl2005software}
Pohl, K., B{\"o}ckle, G., van~der Linden, F.J.: Software Product Line
  Engineering: Foundations, Principles, and Techniques. Springer (2005).
  \doi{10.1007/3-540-28901-1}

\bibitem{sonkin2025workflow}
Sonkin, V., Tudose, C.: Beyond snippet assistance: {A} workflow-centric
  framework for end-to-end {AI}-driven code generation. Computers
  \textbf{14}(3), ~94 (2025). \doi{10.3390/computers14030094}

\bibitem{stackoverflow2025survey}
{Stack Overflow}: Stack overflow developer survey 2025. Online (2025),
  \url{https://survey.stackoverflow.co/2025/}, 49{,}000+ responses from 177
  countries

\bibitem{stumpfle2025llm}
St{\"{u}}mpfle, C., Atray, S., Jazdi, N., Weyrich, M.: Large language model
  assisted transformation of software variants into a software product line.
  In: Proc.\ {ICSR}. pp. 12--20. {IEEE} (2025).
  \doi{10.1109/ICSR66718.2025.00008}

\bibitem{superwhisper2025}
{Superultra Inc.}: Superwhisper: {AI} voice-to-text for {macOS}, {Windows}, and
  {iOS}. \url{https://superwhisper.com/} (2025)

\bibitem{svahnberg2005taxonomy}
Svahnberg, M., van Gurp, J., Bosch, J.: A taxonomy of variability realization
  techniques. Software---Practice and Experience  \textbf{35}(8),  705--754
  (2005). \doi{10.1002/spe.652}

\bibitem{ternava2023specialization}
T{\"e}rnava, X., Acher, M., Combemale, B.: Specialization of run-time
  configuration space at compile-time: An exploratory study. In: SAC SE
  Symposium (2023), hAL Id:
  \href{https://hal.science/hal-03916459v1}{hal-03916459}

\bibitem{ternava2017diversity}
T{\"e}rnava, X., Collet, P.: On the diversity of capturing variability at the
  implementation level. In: Proc.\ {SPLC} (Volume~{B}). pp. 81--88. {ACM}
  (2017). \doi{10.1145/3109729.3109733}

\bibitem{ternava2017tracing}
T{\"e}rnava, X., Collet, P.: Tracing imperfectly modular variability in
  software product line implementation. In: Proc.\ {ICSR}. pp. 112--120.
  Springer (2017). \doi{10.1007/978-3-319-56856-0_8}

\bibitem{ternava2025nullvariability}
T{\"e}rnava, X., Randrianaina, G.A., Lesoil, L., Acher, M.: Small yet
  configurable: Unveiling null variability in software. {HAL} preprint (2025),
  \url{https://hal.science/hal-05097580}, {HAL} Id: hal-05097580

\bibitem{watanabe2026cut}
Watanabe, K., Shirai, T., Kashiwa, Y., Iida, H.: What to cut? {P}redicting
  unnecessary methods in agentic code generation (2026).
  \doi{10.48550/arXiv.2602.17091}

\bibitem{yang2025vibefps}
Yang, P.: Vibe coding a zombie survival {FPS} with {Cursor} + {Sonnet}~3.7 +
  {Superwhisper}. X (Twitter) (Mar 2025),
  \url{https://x.com/petergyang/status/1896793172489155048}

\bibitem{zine2025coevolution}
Zine, M., Quinton, C., Rouvoy, R.: {LLM}-based co-evolution of configurable
  software systems. In: Proc.\ {SPLC}. pp. 27--38. {ACM} (2025).
  \doi{10.1145/3744915.3748460}

\end{thebibliography}
\end{document}